\documentclass[conference]{IEEEtran}
\IEEEoverridecommandlockouts
\usepackage[numbers]{natbib}
\setcitestyle{open={[},close={]}}

\usepackage{amsmath,amssymb,amsfonts}

\usepackage{graphicx}
\usepackage{textcomp}
\usepackage{xcolor}
\usepackage{float}
\usepackage{array}
\usepackage[font=footnotesize,labelfont=bf]{caption}
\usepackage{optidef}
\usepackage{subcaption}
\usepackage[utf8]{inputenc}
\def\BibTeX{{\rm B\kern-.05em{\sc i\kern-.025em b}\kern-.08em
    T\kern-.1667em\lower.7ex\hbox{E}\kern-.125emX}}

\usepackage{algorithm}
\usepackage{algpseudocode}
\usepackage{color, xcolor, colortbl}

\begin{document}

\title{An IoT Blockchain Architecture Using Oracles and Smart Contracts: the Use-Case of a Food Supply Chain\\}
\author{\IEEEauthorblockN{
		Hajar Moudoud \IEEEauthorrefmark{1}\IEEEauthorrefmark{2},
Soumaya Cherkaoui  \IEEEauthorrefmark{1},
Lyes Khoukhi \IEEEauthorrefmark{2}
}

\IEEEauthorblockA{
\IEEEauthorrefmark{1} Department of Electrical and Computer Engineering, Université de Sherbrooke, Canada\\ 
\IEEEauthorrefmark{2} University of Technology of Troyes, France \\ 
\{hajar.moudoud, soumaya.cherkaoui\}@usherbrooke.ca, lyes.khoukhi@utt.fr  }
	


 }

\maketitle
\begin{abstract}
The blockchain is a distributed technology which allows establishing trust among unreliable users who interact and perform transactions with each other. While blockchain technology has been mainly used for crypto-currency, it has emerged as an enabling technology for establishing trust in the realm of the Internet of Things (IoT). Nevertheless, a naive usage of the blockchain for IoT leads to high delays and extensive computational power. In this paper, we propose a blockchain architecture dedicated to being used in a supply chain which comprises different distributed IoT entities. We propose a lightweight consensus for this architecture, called LC4IoT. The consensus is evaluated through extensive simulations. The results show that the proposed consensus uses low computational power, storage capability and latency.
\end{abstract}

\IEEEpeerreviewmaketitle

\section{Introduction}
\label{sec: Introduction}
With the rapid rise of smart devices, smart homes, smart cities, and smart everything, the Internet of Things (IoT) has gained popularity among users \cite{pimrc1}\cite{pimrc2}. IoT can be defined as a group of interconnected things or devices, in a private or a public network, sharing data to provide a service such as automation or monitoring. However, in order to fully take advantage of the IoT paradigm, several problems should be addressed \cite{pimrc3}. These problems include particularly, security and privacy issues, especially when private and business-related information are collected and shared among different entities. Another set of related problems are integrity, trustworthiness, and non-repudiation of the data shared among the different entities \cite{r19}. For example, in a smart hospital, the subject will receive treatment based on data provided by sensors; but could we trust the information provided by these sensors?

Blockchain (BC) has been proposed as a solution to overcome these problems \cite{r20}. Because it is decentralized, BC eliminates the need of having a third party verify data integrity, trustworthiness and non-repudiation. Also, BC does not have a single point of failure. BC is a distributed technology that allows transaction verification by members which could be dishonest. It is an immutable ledger (chain) that maintains a continuously growing set of data records called blocks. Data records store information about each transaction performed by the users. Once a valid block is gathered, it is connected to the last block of the chain.

BC uses cryptography signatures, public/private keys, and a consensus mechanism for appending any new block into the chain. A consensus corresponds to a protocol that establishes an agreement among independent entities about the state of the BC. The ability of BC to reach consensus among dishonest distributed peers provides a high system availability and security for IoT systems involving numerous entities. Still, the computing power needed to run the BC is somehow incompatible with the restrictive features of several IoT systems \cite{r7} \cite{pimrc4}. For example, IoT endpoints generally have limited resources; they are power-constrained with limited computing energy, storage, and bandwidth.

There have been several works that tried to address the challenges of using BC in an IoT context \cite{r22}. However, most of these works, if not all, display shortcomings in one or several of the following aspects; openness, lightweight consensus, use of smart contracts, and Oracles. We define openness as the capacity of non-members of BC to access stored data. For example, when using an architecture such as the one proposed in \cite{r21}, only members of the BC can access data, which can restrict access to information that can be of interest to the public in general. A consensus for IoT BC should be non compute-intensive; in other words, lightweight. This aspect has not been addressed in several works \cite{r23}\cite{r24}. Smart contracts implement a formal model to provide a division of labour between stakeholders, which can be useful for implementing rules and policies. Smart contracts are absent in several works like \cite{r27}. An Oracle is a third-party agent who verifies data that cannot be reached or fetched directly by the BC \cite{r16}; in other terms, data that comes from the physical world (e.g., sensor data). This aspect has not been addressed in many works \cite{r25}\cite{r26}.

In this paper we propose a secure architecture that overcomes the challenges of using BC in an IoT context, that ensures openness, uses a lightweight consensus for IoT (LC4IoT), smart contracts, and Oracles. To accurately illustrate the architecture, a food supply chain use-case will be adopted throughout the paper. Still, the concepts presented in this paper are well suited for other types of IoT applications.

The remainder of this paper is organized as follows. In section II, we present the proposed architecture. Section III details LC4IoT used in the architecture. Section IV studies the performance of LC4IoT in comparison with the consensus used by the Bitcoin architecture. In section V, we give an overview of some proposed BC IoT architectures. Finally, section VI concludes the paper.

\section{FOOD SUPPLY BLOCKCHAIN}
\label{sec: system}
The food supply chain system includes multiple stakeholders such as farmers, distributors, retailers, and consumers (see Fig. 1). These stakeholders can be located in different regions, with produce possibly transiting between several countries before arriving at a consumer. Due to the lack of trust and transparency among tiers, produce tracing becomes challenging. In this context, BC ability to permanently record data and provide real-time access to information could help overcoming the problem of produce traceability. The proposed architecture involves four tiers: overlay IoT network, smart farm, Oracle's network, and Cloud. Thereafter, we define each tier of the architecture.
\subsection{Overlay network}
We introduce a peer-to-peer overlay network including supply chain members. The overlay network forms a distributed network involving multiple stakeholders. This peer-to-peer network is built on top of the supply chain system, enabling stakeholders connected to it to communicate. All members of the supply chain are initialized at the beginning of communication and identified by a public key. A new member is accepted in the overlay network if she is approved by a quorum. A “quorum” is the minimal number of members required to reach an agreement. All members of the overlay network have a list of approved member keys stored in their local storage.

We use smart contracts to ensure that rules and policies are respected by parties in the overlay network. In this architecture we use smart contracts in two ways; first, they are applied by a third party who offers transparency and efficiency, second, they are implemented to govern operations between stakeholders.
\subsection{Smart farm}
The smart farm is comprised of IoT devices, proxy nodes, storage and BC.\\
In general, BC has three main ledger types: public, consortium, and private BC \cite{r1}. In a public BC, everyone can join the network and all BC members are responsible for transaction validation. The consortium BC is different in that, it is partly decentralized; only some members oversee the consensus determination. Generally, consortium BC is built by several organizations. Lastly, private BC restricts access to network members only. Private BC is implemented by a single company or an organization. In the proposed architecture we use a public/private BC. The private BC is used to store private information of the smart farm. The public BC is used for tracking produce, and for providing information to the general public. 
Fig. 2 illustrates the smart farm BC architecture. 

IoT sensors are responsible for gathering data from the field in a smart farm (e.g. RFID tags for cattle). Since IoT devices have limited computing power and energy, we use proxy nodes to outsource computing. The imparting of information between the smart farm IoT devices is referred to as a transaction. Every produce is identified by a public key that changes with every transaction. Data are stored centrally in designated storage, while transactions are recorded in the private/public BC.  
\begin{figure}[t]
	\centering
	\includegraphics[width=6cm, height=6cm]{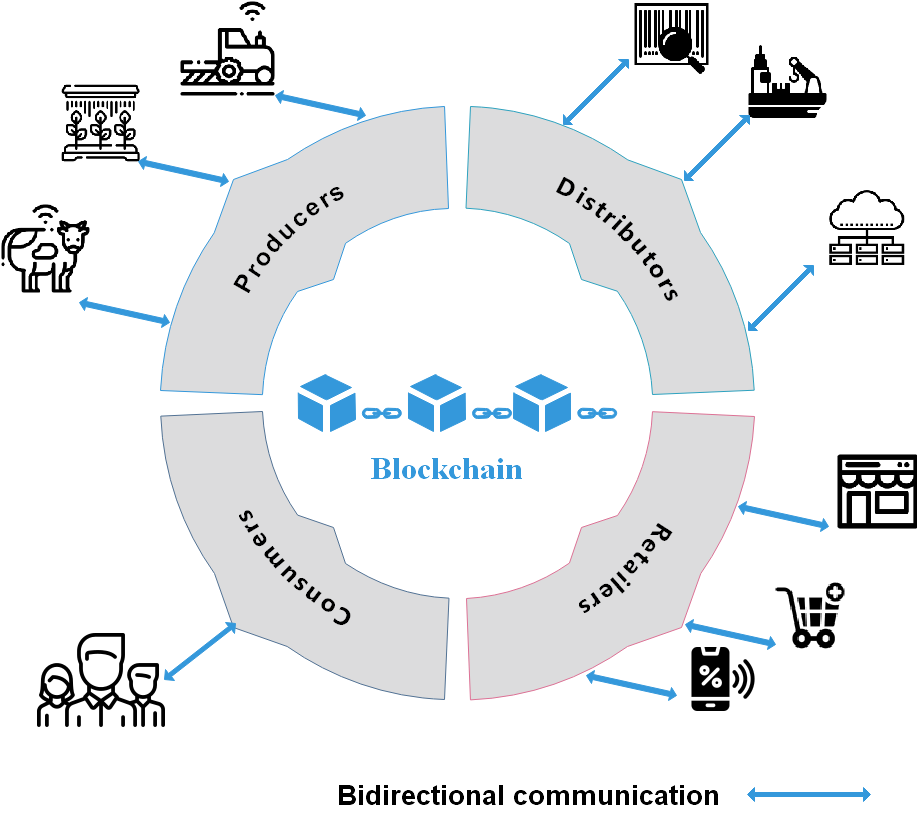}
	\caption{Food supply chain overlay network}
	\label{Overlaynetwork}
\end{figure}
\begin{figure*}[h]
	\centering
	\includegraphics[width=0.9\textwidth ]{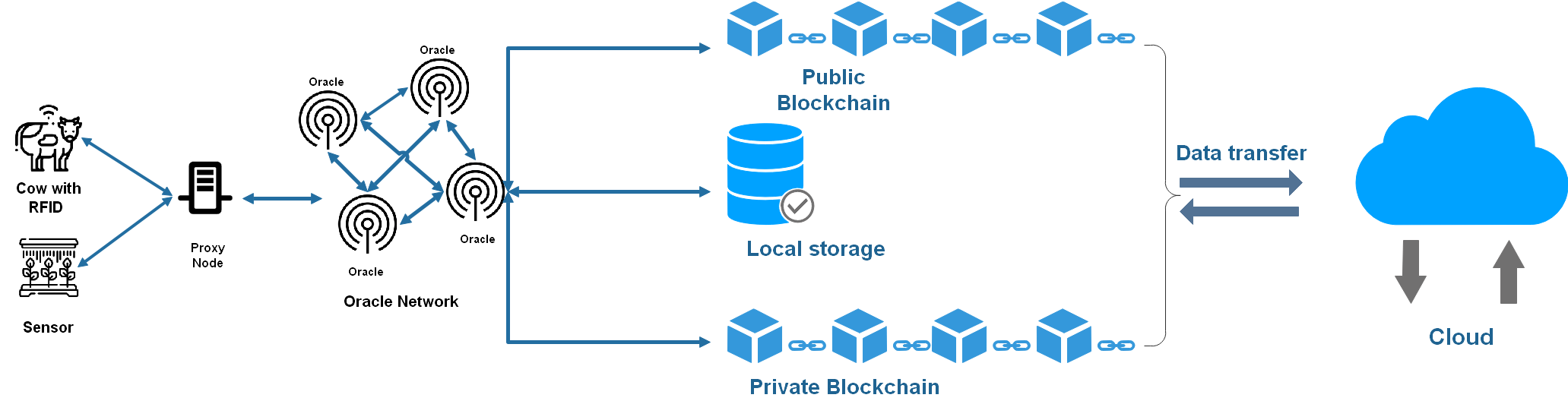}
	\caption{Smart farm architecture based on BC}
	\label{DW}
\end{figure*}
\subsection{Oracle network}
In the food supply chain, data is generally collected from sensors scattered across multiple locations. We use Oracles to check the veracity of sensor data. For example, an Oracle can inform whether the temperature inside a refrigerated track transporting produce has come above a certain threshold during the transportation journey.\\
To verify data, an Oracle $O_i$ needs to compare received data $x$ from a sensor with fetched data  $y$ by the Oracle (see Eq. 1). 
\begin{equation}
O_i(x)=
\begin{cases}
1 \quad \text{if} \, O_i.verify(x,y)==True\\
0 \quad \text{otherwise}. 
\end{cases}
\end{equation}  

In our architecture, multiple Oracles can be used. The Oracles' network is able to divide the approval process for data veracity among multiple parties. 
We define the Oracles' network data verification process as follows: $M$ out of $N$ multi-signature transaction should be reached among Oracle parties. A transaction is valid when $M$ out of $N$ Oracles sign it, that is, the sum is greater than a threshold $\delta$ (Eq.2).\\ 
$\forall O_i, O_i \in \{0,1\}^* \to \{0,1\} ^k, 1 \leq i\leq M, $
\begin{equation}
\delta \leq \sum_{i=1}^{M} O_i(x)
\end{equation}
The $\delta$ value depends on the number of active Oracles and the fault tolerance of the system. Where $M$ is the number of active Oracles in the network and $k$ is a security parameter. 
\subsection{Cloud}  
Cloud stores raw data received from the Oracles' network. Data is either publicly accessible, ensuring data transparency or has limited access to preserve the privacy of stakeholders. Each member of the food supply chain allocates a Cloud space for personal usage, linked with a Cloud public key $CPk$. This process guarantees that data is correctly routed and the source is identified.\\
We propose using a private BC in the Cloud to store data hash, enabling data trustworthiness and non-repudiation. In BC, the hash function is the algorithm used to write a new transaction through the mining process \cite{r14}. It maps data and generates a summary; the unique fingerprint substitutes an input string from arbitrary to a fixed size. The hash enables data integrity by comparing the hash value along with some additional inputs, like timestamp and previous block hash. BC uses the SHA-256 secure hash algorithm, which provides almost a unique and fixed size of 256 bits (32 bytes), requiring low computing power.

\section{LIGHTWEIGHT CONSENSUS}
\label{sec:proposed}
In this section, we explain the transaction verification process and how off-chain storage is performed in the proposed architecture. Also, we detail LC4IoT. 
\subsection{Transaction verification}
In order to append a new transaction in BC, miners need to verify some conditions. This verification process is divided into three steps:
\begin{enumerate}
	\item First, miners verify the sender's signature to validate transaction authenticity. This digital signature authenticates the sender, using the public key stored in the transaction.
	\item Second, miners check if the public key is predefined; which means, the sender public key has a stored transaction in the BC. Otherwise, it is a genesis transaction (see section C).
	\item Third, miners verify the Oracles' network signatures.
\end{enumerate}
If all conditions are validated, data is then transferred to the verified transaction pool for mining.
\subsection{Data transfer off-chain}
The BC could be used as a mediator for data transfer. Generally, there are two methods used to store data within BC; first, data is sent within a transaction, it is the case with Bitcoin \cite{r28}, or, data is stored in smart contracts, it is the case with Ethereum \cite{r29}. Both approaches submit a transaction in the BC. Nevertheless, the BC block size is limited. To solve this problem, several proposals have been put forward. A trivial solution consists in increasing the block size. For example, Bitcoin Cash \cite{r6} upgraded the block size from 1 MB to 8 MB. Still, this solution may affect the operation of nodes and tends to be more expensive. For our architecture, we store raw data off-chain and metadata on-chain, to improve latency and system scalability.

In this section, we outline how data transfer is performed in BC. We explain mining steps in the case where Alice wants to transfer 5GB of data to Bob. Let Alice be a member of the BC, which means that her public key exists in a previous block of the chain. First, Alice will store the 5GB of data in the Cloud. Alice is identified by the triple ($Pk$, $Prk$, $a$) where $Pk$ is the distributed public key, $Prk$ is the secret private key, and $a$ is the address where she stored data. Since block size in BC is limited, Alice will only store metadata on-chain and the actual data off-chain. Alice will share the private key with Bob and encrypt metadata with the public key. Once the transaction is created, Alice will sign it using the private key. A node in the network will verify the state of the transaction, by verifying the signature, public key of both the sender and the receiver. Afterwards, the transaction is sent to the verified transaction pool.

A node in the network can also act as a miner. Miners of the network choose a transaction form the pool and build a block. Different miners could pick the same transaction. Miners in the network will attempt to reach a consensus to append a block. The first miner who reaches a consensus will broadcast the new block to other miners. The transaction becomes a permanent part of the ledger, Bob can now have access to the address where Alice has stored data with her private key. 
\subsection{ Lightweight consensus for IoT (LC4IoT) }
We propose LC4IoT algorithm that integrates the use of Oracles for block appending in the BC. In step 1, nodes in the network fetch a random transaction $ T_k $ from the verified transaction pool and try to create a new block $B_{i+1}$. To simplify, we admit that every block contains only one transaction. Every transaction has the following arguments, $T_k (Pk, CPk, O_i.sig, metadata) $, where $Pk$ is the public key of the data provider, $CPk$ is the Cloud public key, $O_i.sig$ is the Oracles signature and $metadata$ is the metadata that we want to store. The output of the algorithm will be a new block that contains the transaction. In step 2, the algorithm verifies if the stored signature in the transaction belongs to the Oracles' network or not. If the provided signature does not belong to the list of the Oracles' network, the algorithm returns False. Otherwise, the algorithm verifies if the transaction is a storing transaction or a genesis transaction.

To store a transaction, miners should calculate the timestamp $TS$ and the hash of the previous block (step 14). Finally, in steps 15 to 16, the algorithm return the new block $B_{i+1}$, which contains the hash of the previous block, transaction and the hash of the new block.

To ensure the system liveness, which means the growth of the system, we will use PBFT \cite{r18} consensus for genesis transaction. This consensus handles $f$ faulty members of the network. If the sum of stakeholders $ S_i$ in the overlay network is superior to a min $3f+1$ we accept the demand. For a genesis transaction, the system allocates a new public key to the requester, which will be stored in a new block. 
\begin{algorithm}[t]
    \hspace*{\algorithmicindent} \textbf{Input}  {$\forall  T_k \in T$, $\forall B_i \in B$, $\forall O_i \in O$;}\\
    \hspace*{\algorithmicindent} \textbf{Output} {New block $B_{i+1}$};
	\begin{algorithmic}[1]
		\State  \textbf{Initialize:} \\
		$T_k$ : A transaction from the verified transaction pool.\\
		$B_i$ : The last appended block in the blockchain $B$.\\ 
		$O_i$ : Oracle, \\
		$m$ : The number of Oracles in the network\\ 
		$ j=0$
	\State \textbf{while} $ j\leq m || T_k.output(2)\neq O_j.sig$ \textbf{do}\\
		$j \leftarrow j+1$\\
		\State \textbf{end while}
		
		\State  \textbf{if } $ j= m + 1 $\\
		$Return$ $ False$
		\State  \textbf{else}
		\State \quad \textbf{while} $i \geq 0$ $|| $\\
		\quad \quad \quad \quad  $B_i.T_i.outputs(2)==T_k.outputs(o)$ \textbf{do}\\
		\State \quad $ i \leftarrow i-1 $
		\State \quad \textbf{end while}
		\State                      \quad     \textbf{if }$ i= 0 $ \textbf{then}
		\State    \quad \quad    $ T_k \leftarrow GenesisTransaction(T_k) $
		\State \quad \quad update transaction pool 
		\State \quad   \textbf {else}
		\State \quad  $TS \leftarrow date.gettime()$    
		\State \quad  $ B_{i+1}.H \leftarrow$
		$CalH(i1, B_i.H, TS, T_k)$
		\State \quad  $B_{i+1}\leftarrow B_{i+1}( B_i.H, TS, T_k, B_{i+1}.H) $ 
		\State \quad \quad $ Return $ $B_{i+1} $
		\State \quad \textbf{end if }    
		\State \textbf{end if}
		
	\end{algorithmic}
	\caption{Block appending algorithm}
\end{algorithm}
\section{PERFORMANCE ANALYSIS}
\label{sec: evaluation}
In this section, we provide a quantitative performance evaluation of LC4IoT. First, we conduct a simulation to evaluate the computing power and time consumed by the proof of work consensus. Second, we assess LC4IoT: computational power, memory footprint, and latency.
\subsection{Proof of work evaluation}
The proof of work is the consensus algorithm that secures the network by demanding to the requester some work. An example of BC that uses the proof of work consensus is Bitcoin \cite{r17}. To reach an agreement among users, each node of the network calculates a hash value called “Nonce” in the block header. Miners in the network will try to estimate a secret value, then embed it in the block. All information inside the block header will be combined, next inputted to a SHA-256 hash function. The first miner who will reach a hash function output less than a threshold can add the new block to the chain. The new block is then broadcast to network users. In Bitcoin, a valid block hash requires that it starts with several zeros, which refer to difficulty level, adjusted to limit block generations. Currently, the difficulty reached by the Bitcoin is \texttildelow18 zeros \cite{r30}. Since we cannot choose the hash value of a block, miners try several combinations to solve this puzzle. To demonstrate the drawbacks of the proof of work consensus, we implement a proof of work using JavaScript enabling change of difficulty. We simulate the appending of one block on Intel machine Core™ i7-8550U CPU @ 1.80 GHz 2.00 GHz with 16.0 RAM. Fig. 3 shows the results of processing time and computational power. Simulation result proves that we cannot use a proof of work consensus in the context of the food supply chain.
\begin{figure}[t]
	\centering
	\includegraphics[width=8cm, height=4.5cm]{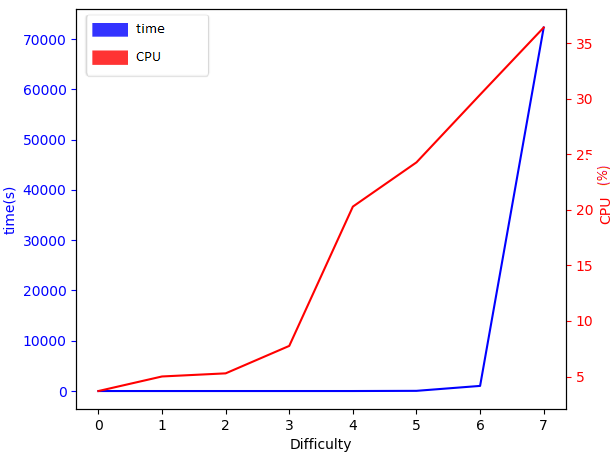}
	\caption{Proof of work evaluation}
	\label{POW}
\end{figure}
\subsection{Performance evaluation}
In this section, we evaluate LC4IoT in contrast to the proof of work consensus. In our experiments, we fixed the difficulty variable to 4 for the proof of work. We use the following evaluation metrics: computational power, memory footprint and processing time. 
We run the simulation 3 times to add 10 blocks into the BC and record CPU evolution during the time. Fig. 4 shows the results of computational power used by both consensuses. $Local$ $max$ annotation refers to the consensus execution peak. 
\begin{figure}[t]
	\centering
	\includegraphics[width=7.5cm,height=9cm,keepaspectratio]{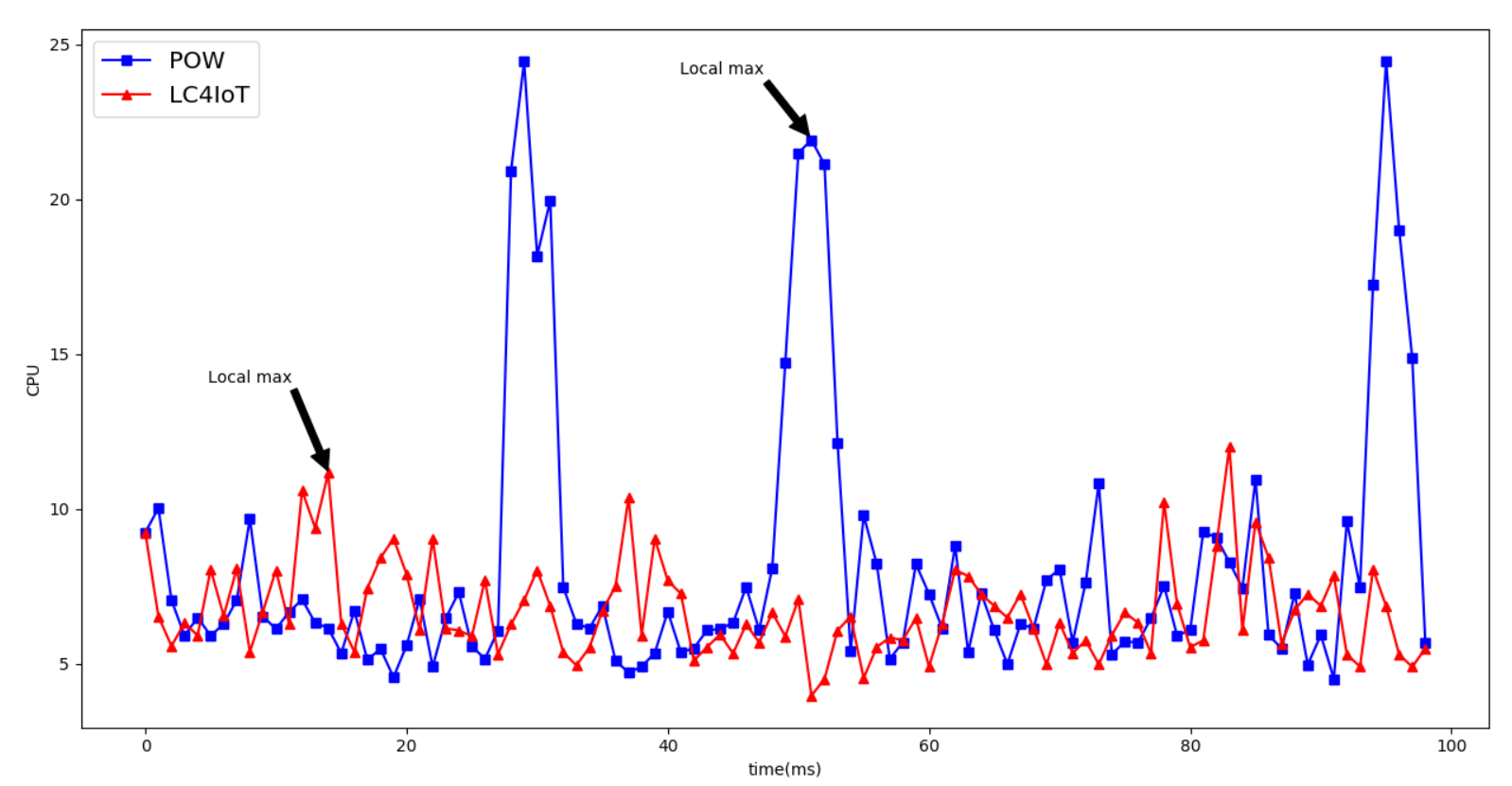}
	\caption {computational power evaluation }
	\label{POW}
\end{figure}
As we mentioned before, we separate transaction verification from block appending, Oracles network oversees transactions, whereas miners append blocks to the ledger. To quantify the advantages of this design decision, we perform simulation while we alter the number of blocks to add to the BC; simulation results are presented in Fig. 5. 

Some BC mining algorithms are memory intensive; RAM footprint required by consensus is higher than system availability. By comparing the memory usage of both consensuses, we display the need for a high-powered computer to perform mining and solve the complex puzzle for the proof of work. Even if proof of work algorithm makes memory requirements seem secondary in comparison with the high-computational power consumed, a lot of RAM usage will make mining run more slowly; leading to an increase in the delay. LC4IoT algorithm improves block verification process and entails a low memory growth rate of $0.362\% $ per block.
The delay metric, shown in Fig. 6, refers to the time consumed by the algorithm to append blocks into the ledger. In Fig. 6 the delay evaluation of the proof of work and LC4IoT are illustrated. As evidenced by the figure, the proof of work delay increases with the number of blocks to add in the BC. The proof of work consumes 7.81s to append 10 blocks, while LC4IoT only needs 0.139s for the same number of blocks.   
\begin{figure}[!h]
	\centering
	\includegraphics[width=8cm,height=6cm,keepaspectratio]{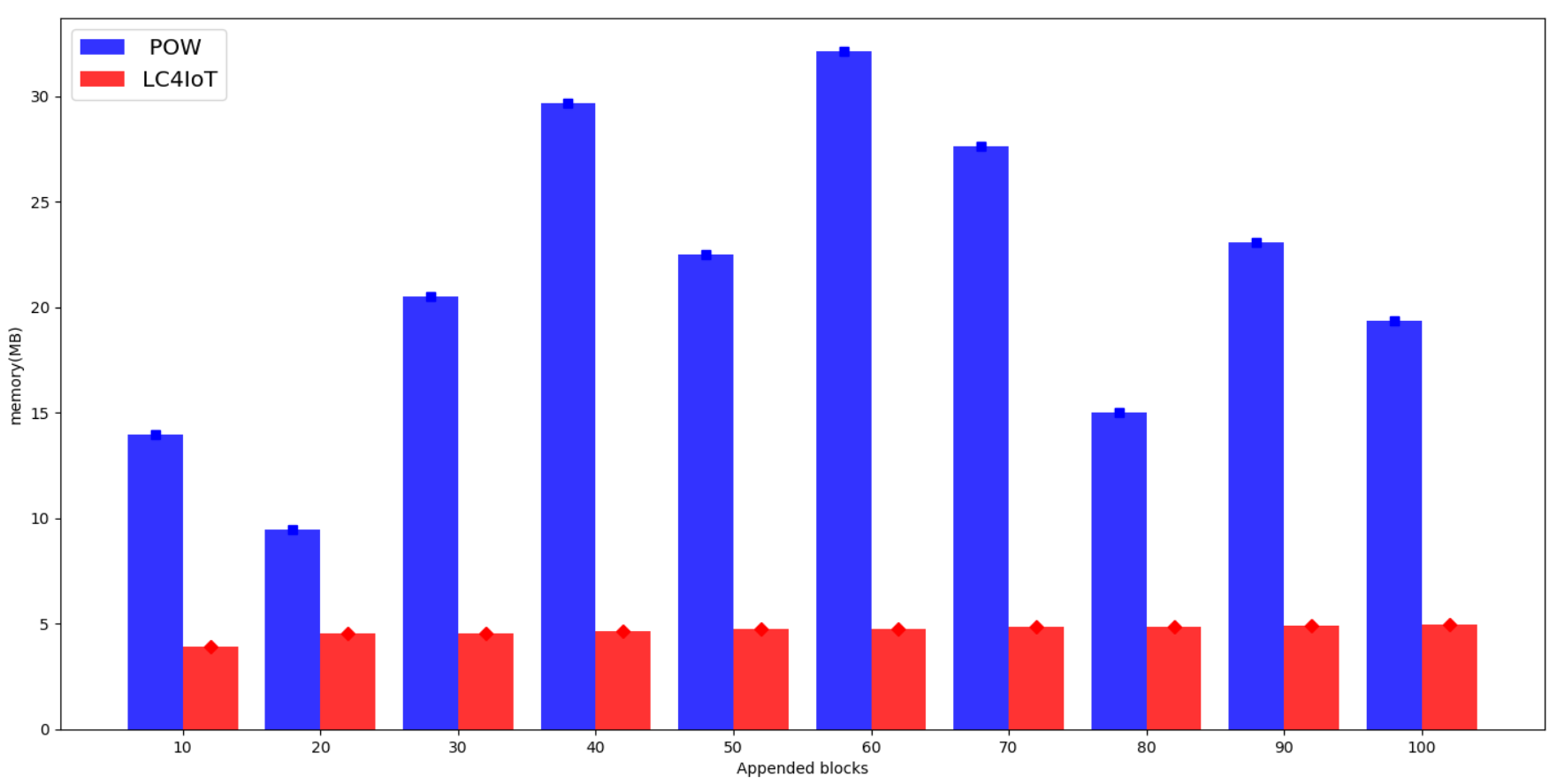}
	\caption {Memory evaluation }
	\label{POW}
\end{figure}
\begin{figure}[!h]
	\centering
	\includegraphics[width=8cm,height=6cm,keepaspectratio]{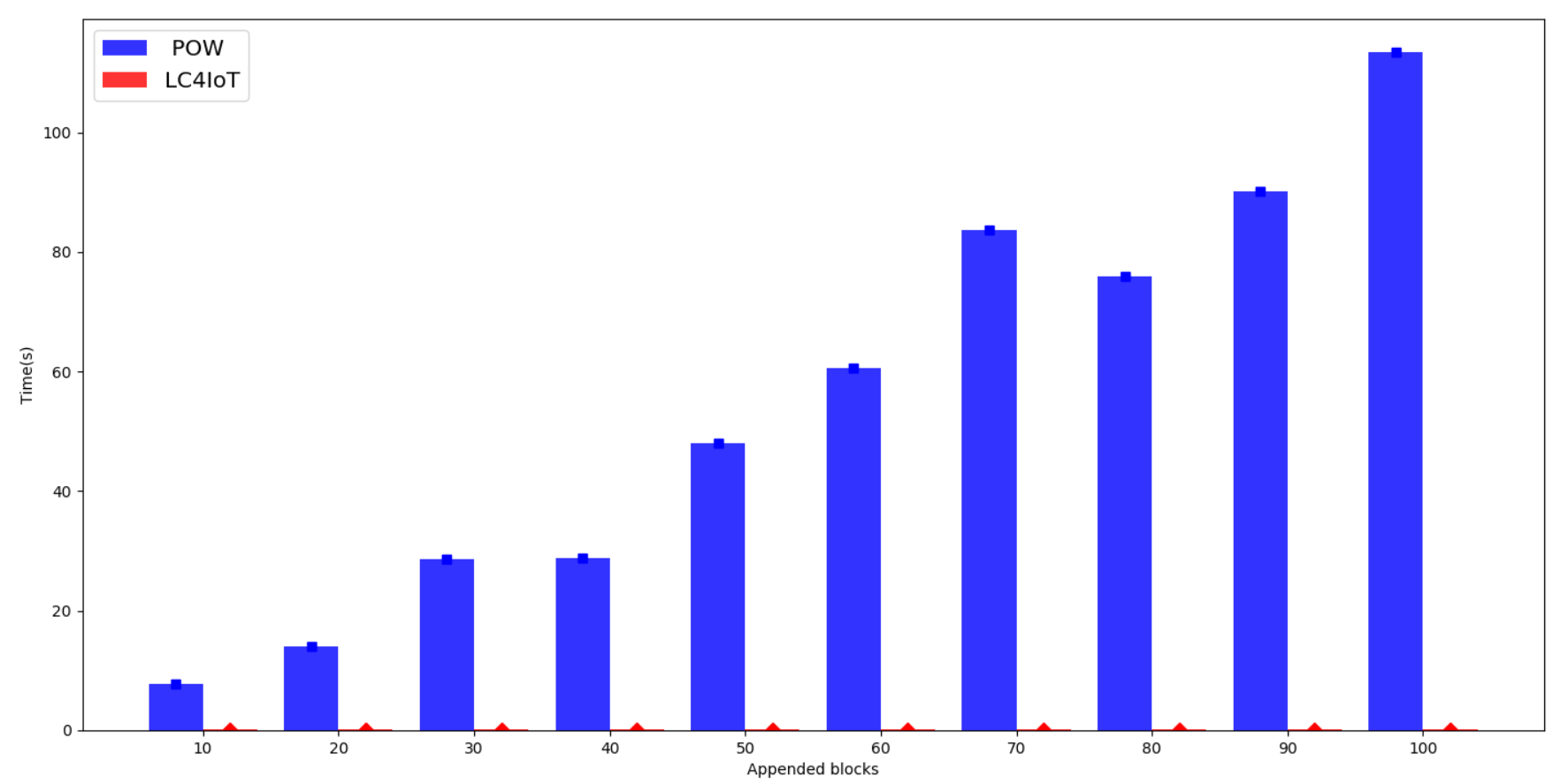}
	\caption {Delay evaluation }
	\label{POW}
\end{figure}

\section{Related Work}
\label{sec: limitations}
Various works have been conducted for using BC in the IoT context. Fernández-Caramés et al. reviewed the state of the art of BC and IoT applications and evaluated the limitations of BC usage for IoT applications \cite{r22}. Likewise, BC who implements smart contracts were used for several purposes. Choi et al. utilized BC smart contracts to secure the control of IoT sensors \cite{r26}. Novo et al. proposed a distributed IoT architecture based on BC smart contracts to ensure access management. This architecture uses a proof of concept consensus \cite{r21} and private BC to secure information. Yet, using a private BC may restrict access to information with a general interest for users. Michelin et al. designed a SpeedyChain framework based on BC, allowing smart vehicles to share data in a secure matter \cite{r27}. Liu et al. presented a BC framework ensuring data integrity \cite{r25}.
Further, BC has helped the supply chain overcoming its limitations such as transparency, reliability and integrity. Tian et al. in \cite{r10} proposed a conceptual framework for agricultural produce traceability using BC and Radio-Frequency Identification (RFID) technology. The BC stores information provided by the RFID technology to help track produce. Korpela et al. in \cite{r11} studied the requirement for supply chain integration with the BC. This investigation proposed the usage of Cloud applications to achieve interoperability in the supply chain. However, both proposals in \cite{r10} and  \cite{r11} do not take into consideration the high number of transactions managed by BC. The supply chain conveys a high volume of communications and requires real-time access to information. A conventional BC needs from seconds to minutes to append a block into the chain. As an example, for Bitcoin BC, block generation takes 10 minutes \cite{r9}. Several BC consensuses require high computational power, which is not suitable for the constrained nature of IoT sensors. In their work, Su et al. developed and implemented a supply chain based on BC technology, using Ethereum consensus \cite{r23}. Pass et al. proposed a fruitChain protocol using the proof of work consensus \cite{r24}. Both the proof of work consensus and the Ethereum consensus need high computing resources for block appending. Dorri et al. proposed a lightweight scalable blockchain (LSB) that reduces the delay for appending data to the BC \cite{r12}. In \cite{r12}, however, the proposed BC architecture does not address how external data outside the network is accessed. 
\section{Conclusion}
\label{sec: Conclusion}
Blockchain (BC) establishes trust among different parties who could be dishonest, enabling data sharing in a secure matter. The use of BC in an IoT context may offer several benefits like trustworthiness and data non-repudiation. However, the constrained nature of IoT sensors is incompatible with the high computational power needed for the BC. In this paper, we presented a secure IoT BC architecture whose usage was illustrated for the use case of a food supply chain. The proposed architecture uses an Oracles' network and smart contracts, ensuring produce traceability and system openness. Further, we proposed using a lightweight consensus for IoT (LC4IoT), which reduces computational power, storage capability, and latency. Simulation results highlighted the effectiveness of the proposed consensus.
\section*{acknowledgement}
\label{sec:acknowledgement}
The authors would like to thank the Natural Sciences and Engineering Research Council of Canada, as well as FEDER and GrandEst Region in France, for the financial support of this research.

\label{Related work}
\bibliographystyle{IEEEtran}
\bibliography{./references}


\end{document}